\documentclass[runningheads]{llncs}
\usepackage[T1]{fontenc}
\usepackage{amsmath,amssymb,amsfonts}
\usepackage{eurosym}
\usepackage{graphicx}
\usepackage{textcomp}
\usepackage{tikz}
\usetikzlibrary{shapes.geometric, arrows.meta, fit, positioning, backgrounds}
\usepackage[table]{xcolor}
\usepackage{colortbl}
\usepackage{orcidlink}
\usepackage{booktabs}
\usepackage{tabularx}
\usepackage{enumitem}
\usepackage{multirow}
\usepackage{array}
\usepackage{makecell}
\usepackage{placeins}
\usepackage{color}
\usepackage{easyReview}
\usepackage{hyperref}

\definecolor{darkgreen}{RGB}{0,120,0}
\begin{document}

\title{A Security-Oriented Lifecycle Model for Large Language Model Systems\thanks{\scriptsize This is a preprint of the following accepted paper: \newline\newline Batzolis, E., Drosatos, G., Katsouros, V., \& Rantos, K. (2026). A Security-Oriented Lifecycle Model for Large Language Model Systems. In P. Kieseberg, F. Skopik, B. Atli, S. Schrittwieser, \& M. Asplund (Eds.), \emph{Availability, Reliability and Security---ARES 2026 EU Projects Symposium Workshops} (Lecture Notes in Computer Science, pp.~1--18). Springer Nature Switzerland.}}
%
%
\author{Eleftherios Batzolis\inst{1,2}\orcidlink{0000-0002-5411-2716} \and
George Drosatos\inst{2}\orcidlink{0000-0002-8130-5775} \and Vassilis Katsouros \inst{2}\orcidlink{0000-0002-4185-2344} \and
Konstantinos Rantos\inst{1}\orcidlink{0000-0003-2453-3904}}
\authorrunning{E. Batzolis et al.}
%
\institute{Department of Informatics, Democritus University of Thrace, Kavala, Greece\\
\email{ebatzoli@cs.duth.gr; krantos@cs.duth.gr}
\and
Institute for Language and Speech Processing, Athena Research Center, Xanthi, Greece\\
\email{e.batzolis@athenarc.gr; gdrosato@athenarc.gr; vsk@athenarc.gr}}
\maketitle             


\begin{abstract}

Large language models are being integrated into critical
infrastructure and enterprise workflows at unprecedented scale,yet the lifecycle frameworks governing their development and operations were designed for operational efficiency rather than security analysis. As a result, security-relevant activities such as data provenance verification, artifact signing, agentic permission control, and decommissioning are often left implicit
or assumed to receive due care. Governance frameworks, in turn, organise requirements around risk levels or management processes without clearly linking them to the lifecycle stages where they apply. This paper addresses both deficiencies. We propose a lifecycle model for LLM systems that supports security analysis by structuring it around security-relevant boundaries rather than
workflow optimisation. The model comprises 32~stages across four core pipeline layers (Data, Model, Distribution, Application), supported by a 12-stage LLMOps pillar and a 9-category governance pillar. Thirteen stages are introduced here as separate units because they expose distinct security concerns that existing frameworks do not clearly distinguish. A governance mapping synthesising the NIST AI~RMF, the EU~AI Act, and ISO/IEC~42001 reveals a structural property of the current regulatory landscape: governance evidence concentrates at deployment-facing
stages, where systems are visible to regulators, while the most consequential decisions, data selection, alignment strategy, and capability boundaries, are made at development-facing stages, where regulatory visibility is lowest.

\end{abstract}

\section{Introduction}
\label{sec:introduction}

Large Language Models (LLMs) have evolved from research artifacts
into foundational components of critical infrastructures and
enterprise workflows~\cite{Zhao_Survey_2025}. Their development
spans multi-stage chains that include data sourcing and curation,
alignment via Reinforcement Learning from Human Feedback
(RLHF)~\cite{Chaudhari_RLHF_2026}, post-training optimisation,
distribution through public model hubs~\cite{Stalnaker_Empirical_2025},
and deployment within orchestrated stacks incorporating
Retrieval-Augmented Generation (RAG), autonomous agents, and
multi-model architectures~\cite{He_Emerged_2026}. Each stage
introduces distinct artifacts, access models, threat actors, and
attack vectors, making the LLM lifecycle a substantially broader
attack surface than that of conventional software systems.

The security research community has responded by cataloguing threat
categories such as data poisoning~\cite{Cina_Wild_2023}, prompt
injection~\cite{Greshake_What_2024}, and model extraction and
privacy leakage~\cite{Hu_Membership_2022}, and by developing
industry frameworks including the OWASP Top~10 for LLM
Applications~\cite{OWASP2025LLMTop10} and MITRE
ATLAS~\cite{MITRE2025ATLAS}. In parallel, governance instruments
such as the NIST AI~RMF~\cite{NIST2023AIRMF} and ISO/IEC~42001~\cite{ISO42001_2023}
have established organisational controls intended to fortify AI
deployments against malicious activity. What remains absent,
however, is a lifecycle structure that connects these perspectives:
a model that decomposes the LLM lifecycle at the boundaries where
security-relevant distinctions arise, thereby enabling systematic
analysis of threats, governance provisions, and operational
requirements at appropriate granularity.

Existing lifecycle and governance frameworks, surveyed in
Section~\ref{sec:background}, exhibit a consistent pattern: stage
boundaries are defined by engineering workflow or by regulatory
logic, neither of which aligns cleanly with security-relevant
distinctions. The result is a four-fold gap. First, a
\emph{granularity gap}: activities associated with different threat
profiles are merged under broader categories, hiding important
security differences. Second, a \emph{security-specificity gap}:
security-relevant activities such as model signing, red teaming,
and input validation are not clearly mapped to distinct lifecycle
stages. Third, an \emph{architectural completeness gap}: no
existing framework combines end-to-end lifecycle coverage with
cross-cutting governance and operational support. Fourth, a
\emph{governance mapping gap}: governance frameworks are not
sufficiently linked to specific lifecycle stages.

We argue that existing LLM lifecycle and governance frameworks are
structurally misaligned for security analysis, and that a
security-oriented lifecycle model, decomposed along attack-surface
boundaries and mapped to governance provisions, makes that
misalignment analysable. We address the four gaps with the
following contributions:

\begin{enumerate}
    \item An \emph{LLM lifecycle model that supports security analysis}, comprising 32~stages across four core pipeline layers. Thirteen are identified here as separate lifecycle stages because they expose distinct security concerns not clearly distinguished in existing frameworks. Several of the underlying activities, such as red teaming and synthetic data
    generation, are already recognised in practice; what is new here is their treatment as separate analytical units for security analysis.

    \item A \emph{12-stage LLMOps operational pillar} synthesised from three academic sources ~\cite{DiazDeArcaya2024LLMOps,Sinha2024LLMOps,Pahune2025TransitioningMLOpsLLMOps} and elevated from a single lifecycle step to a cross-cutting support layer.

    \item A \emph{9-category governance pillar} synthesising the
    NIST AI~RMF~\cite{NIST2023AIRMF}, the EU~AI
    Act~\cite{EUAIAct2024}, and ISO/IEC~42001~\cite{ISO42001_2023},
    that maps governance concerns to lifecycle stages and
    identifies three governance handoffs at layer boundaries.
\end{enumerate}

The remainder of this paper is organized as follows.
Section~\ref{sec:background} reviews related work.
Section~\ref{sec:methodology} describes the methodology.
Section~\ref{sec:core-pipeline} presents the core pipeline.
Sections~\ref{sec:llmops} and~\ref{sec:governance} detail the
LLMOps and governance pillars, respectively.
Section~\ref{sec:comparison} discusses differentiation and
limitations, and Section~\ref{sec:conclusion} concludes.

\section{Related Work}
\label{sec:background}

Existing lifecycle models exhibit a consistent pattern: stage boundaries are typically defined by engineering workflow, which limits their utility for security analysis because activities with different threat profiles are grouped together. MLOps frameworks~\cite{Kreuzberger2023MLOps} provide a general pipeline structure but lack granularity for LLM-specific activities such as alignment or prompt engineering. LLMOps definitions~\cite{DiazDeArcaya2024LLMOps,Sinha2024LLMOps,Pahune2025TransitioningMLOpsLLMOps} address LLM-specific concerns but merge activities with distinct security profiles, treating alignment and fine-tuning as one stage despite different threat actors and poisoning vectors, and combining packaging and distribution across different trust boundaries. Enterprise blueprints~\cite{Chernigovskaya2025StandardizedBPMN} provide operational context but not the stage decomposition needed for per-stage security assessment. These frameworks address the \emph{granularity gap} and \emph{security-specificity gap} only partially.

Security-oriented framings partially address these gaps. Wang et al.~\cite{Wang_SoK_2025} provide the broadest empirical coverage (529~CVEs, 13~stages) but merge distinct threat profiles and lack governance mapping. NIST AI~100-2e2025~\cite{vassilev_adversarial_2025} compresses the upstream pipeline into a single training phase, precluding distinction among annotation, alignment, and distribution stages. ENISA~\cite{european_union_agency_for_cybersecurity_multilayer_2023} and ETSI EN~304~223~\cite{etsi_securing_2025} include decommissioning and apply Confidentiality, Integrity, and Availability (CIA) assessment, validating our design choices, but restrict security analysis to high-level policy guidance without per-stage decomposition. MITRE ATLAS~\cite{MITRE2025ATLAS} provides indispensable threat intelligence but its kill-chain orientation serves threat hunting rather than lifecycle governance mapping. Google's SAIF~\cite{Google2023SAIF} organises security around component-risk mappings and distinguishes Model Creator from Model Consumer roles; its 2.0 release adds agentic concerns that corroborate the need for our Stage~4.11, but its structure remains risk-oriented rather than lifecycle-oriented and does not map to external governance instruments. Supply-chain studies~\cite{wang2024llmsupplychain} validate the four-layer structure as a recurring pattern but predate LLM-specific concerns or provide research agendas without per-stage decomposition. Developer-centric studies~\cite{nguyen_securing_2026} corroborate the security significance of stages our framework foregrounds but organize findings by developer themes rather than lifecycle stages. These works address the \emph{architectural completeness gap} only partially: ENISA and ETSI achieve end-to-end coverage including decommissioning, but none combines this with cross-cutting operational and governance pillars.

On governance, the three dominant frameworks, NIST AI~RMF~\cite{NIST2023AIRMF} (functional axis), the EU~AI Act~\cite{EUAIAct2024} (risk-and-role axis), and ISO/IEC~42001~\cite{ISO42001_2023} (management-cycle axis), embody structurally distinct logics, none of which maps onto the operational stages of LLM development. Organisations must navigate all three, yet none indicates how its requirements relate to specific lifecycle stages, leaving the \emph{governance mapping gap} unaddressed.

The present work responds to these four gaps by proposing a
lifecycle model that combines a core pipeline structured around
security-relevant boundaries with cross-cutting LLMOps and
governance pillars. Table~\ref{tab:comparison} summarises the
comparison across key dimensions, and the differentiation can be
articulated concretely along four axes. On \emph{lifecycle
granularity}, the proposed model decomposes the LLM lifecycle into
four layers and 32~stages, exceeding the resolution of prior
treatments that adopt three architectural
layers~\cite{wang2024llmsupplychain}, thirteen stages without
end-to-end coverage~\cite{Wang_SoK_2025}, or two phases that
collapse the upstream pipeline~\cite{vassilev_adversarial_2025}.
On \emph{security-specific decomposition}, the model identifies
thirteen stages as separate lifecycle units on the basis of
distinct threat actors, exposed artifacts, and access models;
prior frameworks treat these activities either implicitly within
broader stages or as risk categories rather than lifecycle units,
with the partial exception of ENISA~\cite{european_union_agency_for_cybersecurity_multilayer_2023}
and ETSI~\cite{etsi_securing_2025}, which include decommissioning
but stop short of per-stage decomposition. On \emph{end-to-end
coverage}, the model spans data sourcing through decommissioning,
matching ENISA and ETSI in scope while exceeding them in
granularity; most other frameworks, including
SAIF~\cite{Google2023SAIF} and MITRE ATLAS~\cite{MITRE2025ATLAS},
omit decommissioning entirely. On \emph{governance linkage}, the
model provides an explicit provision-to-stage mapping synthesising
the NIST AI~RMF, the EU~AI Act, and ISO/IEC~42001, which no prior
framework in Table~\ref{tab:comparison} offers; ENISA and ETSI
achieve partial governance linkage but do not map provisions to
stages at the resolution required for per-stage accountability
analysis.

\begin{table}[ht]
\centering
\caption{Comparison of LLM/AI Lifecycle and Security Frameworks Against the Four Identified Gaps}
\label{tab:comparison}
\renewcommand{\arraystretch}{1.3}
\scriptsize
\begin{tabularx}{\textwidth}{|p{2.2cm}|X|X|p{1.4cm}|p{1.4cm}|X|}
\hline
\rowcolor[HTML]{F2F2F2}
\textbf{Source} & \textbf{Lifecycle \hspace{1cm}Granularity} & \textbf{Security-Specific Stages} & \textbf{Evidence Base} & \textbf{Gov.\ Linkage} & \textbf{End-to-End Coverage} \\
\hline
Wang et al.~\cite{wang2024llmsupplychain}
& 3 layers (infra, model, app)
& \textcolor{red}{No}
& Theoretical
& None
& \textcolor{red}{No} (no decommissioning stage) \\
\hline
Steidl et al.~\cite{steidl2023pipeline}
& 4 stages (general AI)
& \textcolor{red}{No}
& Mixed
& None
& \textcolor{red}{No} (no decommissioning stage) \\
\hline
Wang et al.~\cite{Wang_SoK_2025}
& 13 stages, 3 layers
& Partial
& Empirical
& None
& \textcolor{red}{No} (no decommissioning stage) \\
\hline
NIST AI 100-2e2025~\cite{vassilev_adversarial_2025}
& 2 phases
& \textcolor{red}{No}
& Theoretical
& None
& \textcolor{red}{No} (no decommissioning stage) \\
\hline
Patel et al.~\cite{patel_towards_2025}
& 3 attack families
& \textcolor{red}{No}
& Mixed
& None
& \textcolor{red}{No} \\
\hline
Huang et al.~\cite{huang_lifting_2024}
& Stakeholder view
& \textcolor{red}{No}
& Mixed
& None
& \textcolor{red}{No} \\
\hline
Nguyen et al.~\cite{nguyen_securing_2026}
& Developer themes
& \textcolor{red}{No}
& Empirical
& None
& \textcolor{red}{No} \\
\hline
Jiang et al.~\cite{Jiang_Empirical_2022}
& Issue categories
& \textcolor{red}{No}
& Empirical
& None
& \textcolor{red}{No} \\
\hline
ENISA~\cite{european_union_agency_for_cybersecurity_multilayer_2023}
& Concept to decom.
& Partial
& Theoretical
& Partial
& \textcolor{darkgreen}{Yes} \\
\hline
ETSI EN 304 223~\cite{etsi_securing_2025}
& 5 phases, 13 princ.
& Partial
& Theoretical
& Partial
& \textcolor{darkgreen}{Yes} \\
\hline
MITRE ATLAS~\cite{MITRE2025ATLAS}
& Kill chain tactics
& Partial
& Mixed
& None
& \textcolor{red}{No} \\
\hline
Google SAIF~\cite{Google2023SAIF}
& Risk-oriented (15 risks)
& Partial
& Mixed
& None
& \textcolor{red}{No} (no decommissioning stage) \\
\hline
\rowcolor[HTML]{F2F2F2}
\textbf{This work}
& \textbf{4 layers, 32 stages}
& \textbf{\textcolor{darkgreen}{Yes} (13 stages)}
& \textbf{Mixed}
& \textbf{Explicit}
& \textbf{\textcolor{darkgreen}{Yes}} \\
\hline
\end{tabularx}
\vspace{1mm}
\parbox{\textwidth}{\scriptsize Columns map to the four gaps from Section~\ref{sec:introduction}: Lifecycle Granularity, Security-Specific Stages, End-to-End Coverage, and Governance Linkage. Evidence Base indicates each work's empirical grounding.}
\end{table}



\section{Methodology}
\label{sec:methodology}

The proposed framework is analytically derived rather than empirically measured: it is constructed through structured synthesis of prior lifecycle, security, and governance literature, and provides a scaffold against which empirical security data can subsequently be organised rather than reporting new empirical findings.

The lifecycle model was constructed through analytical synthesis of three source categories: (i)~existing lifecycle frameworks (Section~\ref{sec:background}), which provided the baseline stage inventory; (ii)~the supply-chain and security literature \cite{wang2024llmsupplychain,Gokkaya_Software_2025,Wang_SoK_2025} \cite{Cina_Wild_2023,Greshake_What_2024,Hu_Membership_2022}, which documented threats at stages existing frameworks omit; and (iii)~governance instruments~\cite{NIST2023AIRMF,NIST2024GenAIProfile,EUAIAct2024,ISO42001_2023}, whose requirements imply lifecycle stages that operational frameworks do not make explicit.

The organising principle is \emph{attack surface exposure},
characterised along three dimensions: the threat actors with access to a stage, the artifacts exposed at that stage, and the access model governing how those artifacts are reached. Stage boundaries are placed where one or more of these dimensions change in a way that materially alters the security concerns applicable to the activity in question. We distinguish four threat actor classes along this principle. \emph{External adversaries} (e.g., data poisoners, supply-chain attackers) operate without legitimate access, acting through indirect channels such as publicly scraped data or compromised upstream dependencies. \emph{Insiders} (e.g., compromised annotators, preference evaluators) hold privileged access to internal development artifacts and corrupt training or alignment signals at their point of creation. \emph{Supply-chain intermediaries}, further divided into infrastructure operators (e.g., model hub platforms) and dependency maintainers (e.g., library authors), control artifacts in transit between development and deployment. \emph{End users} (e.g., prompt injectors, agent manipulators) hold legitimate credentials and operate within sanctioned interfaces but may abuse those
interfaces in ways that external adversaries cannot, because end users interact with the system after deployment-time controls have already been applied; the mitigations applicable to this class (input validation, output filtering, permission control) are correspondingly distinct from those applicable to external adversaries (perimeter and supply-chain controls). The \emph{access model}, namely the type of artifact exposed, the level of privilege available, and the authentication boundary in place, is the third dimension of attack surface exposure and the dimension along which threat actor classes are distinguished in practice. Lifecycle operations on data, models, or deployed systems are grouped into the same stage when they share similar threat actors, exposed artifacts, and access patterns; they are separated when these characteristics differ enough to warrant independent security analysis. This principle operates as a guiding heuristic rather
than a formal decision procedure.

As a worked example, the separation of Supervised Fine-Tuning (SFT, Stage 2.3) from Alignment (Stage~2.4), here understood as preference-based fine-tuning that shapes model behaviour toward human preferences and safety
constraints~\cite{Chaudhari_RLHF_2026}, follows directly from this principle. SFT poisoning is mounted by data curators against instruction-response pairs through instruction injection, whereas alignment poisoning is mounted by preference evaluators and reward-model trainers against preference rankings and reward models, through label corruption~\cite{Kusaka_CostMinimized_2025} and reward-model backdoors~\cite{Wang_RLHFPoison_2024}. The three dimensions of attack surface exposure thus differ across the two activities, warranting separate treatment. The same reasoning, applied across the lifecycle, separates other commonly merged pairs (data
sourcing from annotation, packaging from signing) and underlies the introduction of thirteen stages whose security concerns are not represented as separate units in existing lifecycle models.

The governance pillar synthesizes provisions from the NIST
AI~RMF~\cite{NIST2023AIRMF} (including AI~600-1~\cite{NIST2024GenAIProfile}), the EU~AI Act~\cite{EUAIAct2024}, and ISO/IEC~42001~\cite{ISO42001_2023}, selected because they represent three complementary approaches: voluntary guidance, prescriptive regulation, and certifiable standardization. Provisions were inventoried, assigned to the lifecycle stage(s) where they are most relevant, and aggregated into nine categories (defined in Section~\ref{sec:gov-categories}). Governance handoffs, that is, transfers of responsibility and accountability, were identified at points where responsibility shifts, where specific evidence must be passed between roles, or where decisions affect later stages of the lifecycle. The structural asymmetry was identified by classifying each provision as either evidence-producing (e.g., documentation, logging, reporting) or decision-shaping
(e.g., data selection, alignment strategy, capability
boundaries). This classification revealed that
evidence-producing provisions concentrate at Layers~3--4, while decision-shaping provisions concentrate at Layers~1--2.

The LLMOps pillar was derived through three-phase synthesis of three primary
sources~\cite{DiazDeArcaya2024LLMOps,Sinha2024LLMOps,Pahune2025TransitioningMLOpsLLMOps}: consensus identification (six stages appearing in at least two sources), elevation of implicitly treated concerns to named stages (Guardrails and Content Safety, Security and Compliance Automation), and gap closure (four stages absent from the initial inventory). The twelve stages were cross-referenced
against practitioner sources~\cite{Chernigovskaya2025StandardizedBPMN};
Table~\ref{tab:llmops-sources} shows the derivation provenance.


\section{Core Pipeline: Data, Model, Distribution, Application}
\label{sec:core-pipeline}

\begin{figure}[ht]
\centering
\includegraphics[width=\textwidth]{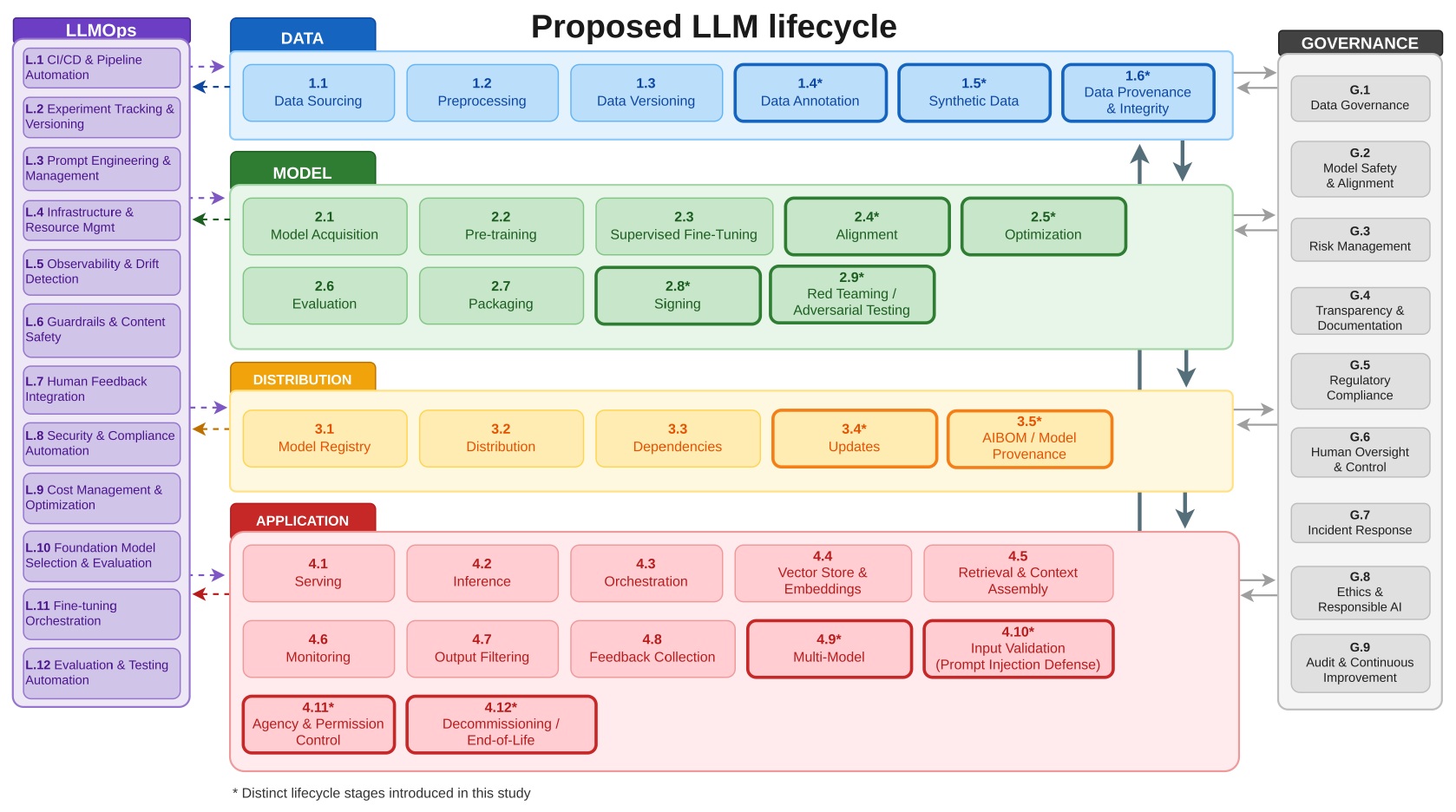}
\caption{LLM Lifecycle: four core pipeline layers (Data, Model, Distribution, Application), the LLMOps operational pillar (left), the Governance pillar (right), and the continuous feedback loop.}
\label{fig:lifecycle}
\end{figure}

The lifecycle model comprises four core layers arranged as a sequential pipeline: Data, Model, Distribution, and Application, comprising 32~stages in total, and is supported by two cross-cutting pillars: the LLMOps pillar (Section~\ref{sec:llmops}) and the Governance pillar (Section~\ref{sec:governance}). A continuous feedback loop connects the Application layer back to the Data layer; this loop has security implications, as adversarial feedback collected at Layer~4 can be reintroduced into training data at Layer~1, propagating issues across iterations. Figure~\ref{fig:lifecycle} presents the complete model. The subsections below define each layer, focusing on the 13~additional stages; stages that refine existing categories are summarised in Figure~\ref{fig:lifecycle}. Each stage is exposed to different types of threats, and representative examples are provided to illustrate these differences.

\subsection{Layer~1: Data (Stages 1.1--1.6)}
\label{sec:pipeline-data}

The Data layer produces, curates, and governs the corpora from which an LLM acquires its capabilities; a compromise at any stage \add{of this layer }propagates through every subsequent layer~\cite{Cina_Wild_2023}. We decompose the Data layer into six stages, two more than typical MLOps frameworks~\cite{Kreuzberger2023MLOps} and three more than NIST's monolithic training-data category~\cite{vassilev_adversarial_2025}, reflecting differences in security-relevant concerns. Stages~1.1--1.3 (Data Sourcing, Preprocessing, Versioning) refine existing categories. Three additional stages address gaps: \emph{Data Annotation}~(1.4) introduces a human-in-the-loop attack surface (label manipulation~\cite{Aryal_Analysis_2022}, annotator data exposure, platform injection~\cite{Saeeda_RE_2025}) materially distinct from automated stages; \emph{Synthetic Data}~(1.5) isolates risks unique to model-generated corpora, including model collapse through recursive training~\cite{Shumailov_AI_2024} and bias amplification through feedback loops; and \emph{Data Provenance and Integrity}~(1.6) addresses the metadata infrastructure recording data origins and transformations, which constitutes a bidirectional attack surface: provenance records can be tampered with to disguise poisoned data~\cite{Baracaldo_Mitigating_2017}, while overly detailed records may leak sensitive infrastructure information~\cite{Spoczynski_Atlas_2025}.

\subsection{Layer~2: Model (Stages 2.1--2.9)}
\label{sec:pipeline-model}

The Model layer spans all activities transforming curated data into a deployable artifact. Its nine-stage decomposition substantially extends prior two- or three-stage treatments~\cite{DiazDeArcaya2024LLMOps}. Therefore, we separate acquisition from training (supply-chain vs infrastructure trust), SFT from alignment (different threat actors and poisoning vectors), and introduce dedicated stages for packaging, signing, and adversarial testing, as they involve different security concerns. Stages~2.1--2.3 (Acquisition, Pre-training, SFT), 2.6~(Evaluation), and 2.7~(Packaging) refine existing categories. Four additional stages, introduced here, address gaps: \emph{Alignment}~(2.4) is important for model safety, as preference data and reward models can be poisoned to embed backdoors~\cite{Wu_Preference_2025}, and issues introduced during alignment can weaken safety mechanisms used in later stages. \emph{Optimization}~(2.5) addresses post-training transformations (quantization, pruning, merging) where models may appear benign at full precision but behave differently when quantized~\cite{Egashira_Exploiting_2024}, and compromised models propagate backdoors through weight averaging~\cite{Zhang_Badmerging_2024}. \emph{Model Signing}~(2.8) addresses cryptographic attestation of artifacts, architecturally distinct from both packaging and distribution, where signing key exposure~\cite{Sood_Malicious_2025} and the absence of widely adopted standards leave many models distributed without integrity verification. \emph{Red Teaming}~(2.9) is separated from general evaluation because it measures robustness under attack rather than capability; red-team findings can also be misused~\cite{OWASP_Top_2025}, and models can be crafted to evade safety evaluations while remaining harmful in production~\cite{Zhao_Survey_2025}.
Together, these stages make the Model layer the locus of the most consequential and least reversible security decisions: compromises introduced here, whether through poisoned alignment data, backdoored reward models, or absent integrity attestation, propagate through every downstream layer and are seldom remediable by deployment-time controls alone.

\subsection{Layer~3: Distribution (Stages 3.1--3.5)}
\label{sec:pipeline-dist}

The Distribution layer covers all activities between production of a signed artifact and its instantiation in a running application, which is a major supply-chain attack surface~\cite{Gokkaya_Software_2025}. Stages~3.1--3.3 (Registry, Distribution, Dependencies) refine existing categories. Two additional stages, introduced here, address gaps: \emph{Updates}~(3.4) separates post-deployment refresh mechanisms from initial distribution because the trust model differs: updates propagate automatically, often without manual review, and malicious payloads can be delivered through established channels. \emph{AI Bill of Materials (AIBOM) and Model Provenance}~(3.5) addresses generation and verification of AIBOMs~\cite{OWASP2025AIBOM}, where AIBOM records may be forged~\cite{Vandendriessche_AIBoMGen_2026} while overly detailed AIBOMs may disclose proprietary details enabling targeted attacks~\cite{Spoczynski_Atlas_2025}.

\subsection{Layer~4: Application (Stages 4.1--4.12)}
\label{sec:pipeline-app}

The Application layer spans model instantiation through runtime operation, monitoring, and eventual decommissioning: it is the largest layer reflecting the diverse architectural patterns characterising modern LLM deployments. Stages~4.1--4.8 (Serving, Inference, Orchestration, Vector Store, Retrieval, Monitoring, Output Filtering, Feedback Collection) are present in varying degrees across existing frameworks. Serving~(4.1) and Inference~(4.2), while sharing end users as a threat actor class, differ in artifact type (infrastructure configurations versus inference requests and outputs) and attack vector space (resource exhaustion and configuration tampering versus prompt injection and output manipulation), justifying their separation. Four additional stages, introduced here, address gaps: \emph{Multi-Model}~(4.9) captures interaction effects that are not present in single-model deployments, including cross-tenant information channels through shared KV-cache entries that allow one tenant's context to leak into another's generation~\cite{Wu_Know_2025} and adversarial prompt contagion across multi-agent systems~\cite{Ding_MoEcho_2025}. \emph{Input Validation}~(4.10) separates defensive infrastructure, such as prompt guards and classification models, from inference, as encoding manipulation and tokenizer-level tricks bypass detection~\cite{MetaBreak_Jailbreaking_2026} and probing attacks reverse-engineer filter rules. \emph{Agency and Permission Control}~(4.11) governs what autonomous agents are permitted to do~\cite{OWASP_Top_2025}, encompassing tool-use authorization protocols such as the Model Context Protocol (MCP) and Agent-to-Agent (A2A) that define the interface through which agents invoke external capabilities: agents with overly broad permissions can be manipulated into unauthorized actions~\cite{OperantAI_Shadow_2025}, privilege escalation exploits inter-agent trust~\cite{Hou_Model_2026}, and individually permitted tool invocations can be chained to achieve unauthorized outcomes. \emph{Decommissioning}~(4.12) addresses model retirement, which is treated explicitly only by a small number of the frameworks reviewed in Table~\ref{tab:comparison}: deprecated weights may contain extractable data, orphaned API keys may remain valid, and namespaces of retired models can be claimed by adversaries.

\section{The LLMOps Pillar}
\label{sec:llmops}

LLMOps should not be treated as a single step within the Model layer. Prior work~\cite{DiazDeArcaya2024LLMOps,Sinha2024LLMOps,Pahune2025TransitioningMLOpsLLMOps} characterises LLMOps as an end-to-end orchestration concern, and the compromise of LLMOps infrastructure carries cascading consequences that exceed those of any single pipeline stage, since it can simultaneously disable enforcement mechanisms and suppress detection. We therefore elevate LLMOps to a cross-cutting pillar comprising twelve \add{sub-}stages\footnote{We use \emph{stage} for core pipeline units and \emph{sub-stage} for LLMOps units throughout.} derived through the three-phase process described in Section~\ref{sec:methodology}. We define a sub-stage as \emph{consensus} when it appears as a named function in at least two of the three primary sources, as \emph{elevated} when it is treated implicitly in one or more sources but never named as a distinct operational function, and as \emph{gap-closure} when it is absent from the inventory of all three primary sources. Under
this criterion, seven sub-stages are consensus, two are elevated, and three are gap-closure. Together, these sub-stages capture the main operational functions required to manage LLM systems across the lifecycle, and make explicit several aspects that are treated only implicitly in existing definitions. Table~\ref{tab:llmops-sources} summarises how each sub-stage relates to existing definitions of LLMOps and indicates the synthesis phase from which it was derived.

\begin{table}[ht]
\centering
\caption{LLMOps Sub-stage Derivation: Source Coverage and Phase}
\label{tab:llmops-sources}
\renewcommand{\arraystretch}{1.15}
\scriptsize
\begin{tabular}{|p{4.6cm}|c|c|c|l|}
\hline
\rowcolor[HTML]{F2F2F2}
\textbf{Sub-stage} & \textbf{D\textsuperscript{a}} & \textbf{S\textsuperscript{b}} & \textbf{P\textsuperscript{c}} & \textbf{Phase} \\
\hline
L.1~~CI/CD \& Pipeline Automation          & \checkmark & \checkmark & \checkmark & Consensus \\
L.2~~Experiment Tracking \& Versioning     & \checkmark & \checkmark &            & Consensus \\
L.3~~Prompt Engineering \& Management      & \checkmark & \checkmark & \checkmark & Consensus \\
L.4~~Infrastructure \& Resource Mgmt       & \checkmark &            & \checkmark & Consensus \\
L.5~~Observability \& Drift Detection      & \checkmark & \checkmark & \checkmark & Consensus \\
L.6~~Guardrails \& Content Safety          &            & $\sim$     & $\sim$     & Elevation \\
L.7~~Human Feedback Integration            & \checkmark & \checkmark &            & Consensus \\
L.8~~Security \& Compliance Automation     & $\sim$     &            &            & Elevation \\
L.9~~Cost Management \& Optimization       & \checkmark & \checkmark & \checkmark & Consensus \\
L.10~Foundation Model Selection \& Eval.   &            & \checkmark &            & Gap closure \\
L.11~Fine-tuning Orchestration             &            & \checkmark &            & Gap closure \\
L.12~Evaluation \& Testing Automation      &            & \checkmark &            & Gap closure \\
\hline
\multicolumn{5}{c}{\scriptsize
\textsuperscript{a}Diaz-De-Arcaya~\cite{DiazDeArcaya2024LLMOps};
\textsuperscript{b}Sinha~\cite{Sinha2024LLMOps};
\textsuperscript{c}Pahune~\cite{Pahune2025TransitioningMLOpsLLMOps};
\checkmark\,=\,named function; $\sim$\,=\,implicit treatment.}
\end{tabular}
\end{table}
\FloatBarrier

Every named function in Sinha et al.~\cite{Sinha2024LLMOps} is represented among the twelve sub-stages, with two exceptions, Infrastructure and Resource Management and Security and Compliance
Automation, that are present in other primary sources but not in Sinha, which together suggests that the principal operational functions of LLMOps are accounted for. The relationship between the LLMOps pillar and the core pipeline is bidirectional: in the forward direction, LLMOps orchestrates stage progression, provisions resources, enforces safety constraints, and applies
validation checks; in the reverse direction, the pipeline generates the signals, dataset metadata, training metrics, user interactions, and inference costs, on which LLMOps operates. 
This bidirectional relationship has direct security implications, since controls are applied through LLMOps while security events originate from the pipeline, with the consequence that a single compromise of LLMOps infrastructure can simultaneously degrade enforcement and suppress detection. 
Consider, for instance, an attacker who compromises CI/CD and Pipeline Automation~(L.1): such
an attacker may introduce a poisoned model into the deployment workflow, tamper with experiment-tracking and versioning records~(L.2) so that the poisoned model appears to pass evaluation, disable the guardrail configurations~(L.6) that would otherwise catch anomalous outputs, and suppress the observability alerts~(L.5) that would notify defenders. 
The resulting cascade illustrates how a single entry point can propagate across four
LLMOps sub-stages, a pattern that motivates treating LLMOps as a distinct security-analytical layer rather than as a single operational concern.


\section{The Governance Pillar}
\label{sec:governance}

\begin{figure}[ht]
\centering
\includegraphics[width=\linewidth]{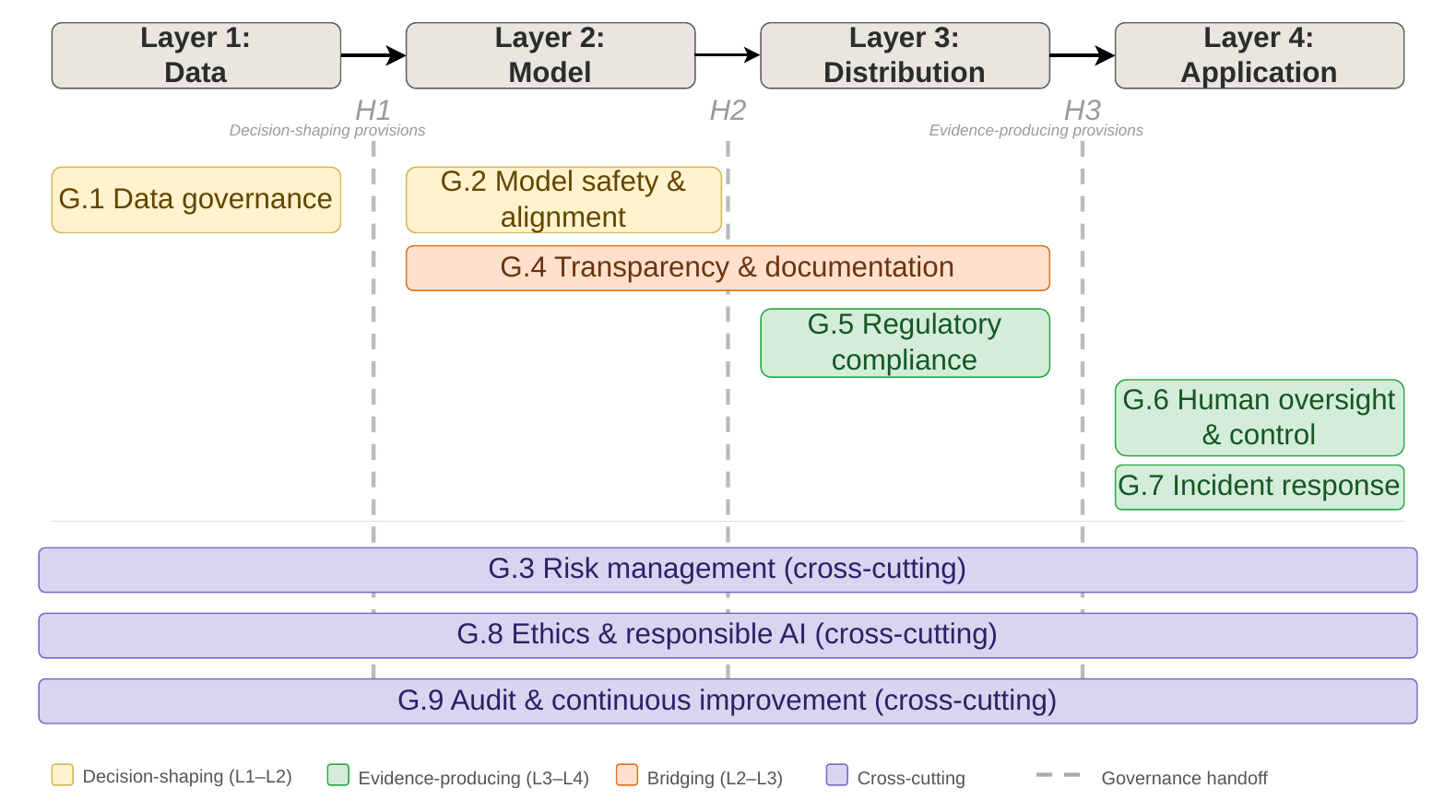}
\caption{Governance stack mapped to the LLM lifecycle: nine
categories assigned to their primary lifecycle layers.}
\label{fig:llm_lifecycle_governance_stack_v2}
\end{figure}

The governance pillar synthesises the NIST AI~RMF~\cite{NIST2023AIRMF},
the EU~AI Act~\cite{EUAIAct2024}, and ISO/IEC~42001~\cite{ISO42001_2023}
into a unified, lifecycle-mapped governance structure.
Figure~\ref{fig:llm_lifecycle_governance_stack_v2} presents the
mapping of the governance stack to the LLM lifecycle layers.

\subsection{Design Rationale and Nine Categories}
\label{sec:gov-categories}

The three frameworks embody structurally distinct logics:
functional (NIST), risk-and-role (EU~AI Act), and management-cycle
(ISO/IEC~42001), none of which maps directly onto development
stages. The governance pillar reframes provisions along a lifecycle
axis by assigning each provision to the lifecycle stage at which
the activities it regulates actually take place. The EU~AI Act, in
particular, imposes obligations on providers of \emph{General-Purpose
AI (GPAI)} models, that is, foundation models offered for
downstream integration; these obligations (Arts.~53--55) are
treated here as a distinct sub-class of provisions within the
existing categories rather than as a separate governance framework.
The nine categories, ordered from data-proximal to cross-cutting,
are: Data Governance~(0.1), primarily Layer~1; Model Safety and
Alignment~(0.2), primarily Layer~2; Risk Management~(0.3),
cross-cutting; Transparency and Documentation~(0.4), concentrated
at the Layer~2/3 boundary; Regulatory Compliance~(0.5), primarily
Layer~3, carrying the heaviest enforcement weight (EU~AI Act fines
up to  \EUR15~million or 3\% of global turnover); Human Oversight
and Control~(0.6), primarily Layer~4; Incident Response~(0.7),
primarily Layer~4; Ethics and Responsible AI~(0.8), cross-cutting;
and Audit and Continuous Improvement~(0.9), cross-cutting.
Table~\ref{tab:gov-mapping} presents a representative excerpt of
the provision-to-stage mapping.

\begin{table}[htb]
\centering
\caption{Governance Provision-to-Stage Mapping}
\label{tab:gov-mapping}
\renewcommand{\arraystretch}{1.2}
\scriptsize
\begin{tabular}{|p{2.65cm}|p{0.8cm}|p{8.3cm}|}
\hline
\rowcolor[HTML]{F2F2F2}
\textbf{Category} & \textbf{Layer} & \textbf{Key Provisions} \\
\hline
0.1 Data Gov.
& L1
& EU AI Act Art.~10: data governance and quality; GPAI Art.~53(1)(d): training data summary; N:~MAP~1, MAP~4: data sources and third-party risks; ISO A.4: data resource controls \\
\hline
0.2 Model Safety
& L2
& EU AI Act Art.~9: risk management; GPAI Art.~55: adversarial testing; N:~MEASURE~1--2: evaluation and risk assessment; GenAI~600-1: hallucination and bias; ISO Cl.~6.1: risk and impact assessment \\
\hline
0.4 Transp.\ \& Doc.
& L2/L3
& EU AI Act Art.~11: technical documentation; GPAI Art.~53: transparency report; N:~GOVERN~4: transparency culture; ISO Cl.~7.5: model cards and AIMS records \\
\hline
0.5 Reg.\ Compl.
& L3
& EU AI Act Art.~43: conformity assessment; Arts.~16--17: provider QMS; N:~MANAGE~1, GOVERN~1.1: risk treatment and legal requirements; ISO Cl.~8.1: release controls; A.10.3: supplier oversight \\
\hline
0.6 Human Oversight
& L4
& EU AI Act Art.~14: human oversight; Art.~50: AI disclosure; Art.~27: FRIA; N:~GOVERN~5: stakeholder engagement; ISO A.9.3: oversight controls \\
\hline
0.7 Incident Resp.
& L4
& EU AI Act Art.~12: logging; Arts.~72--73: post-market monitoring and incident reporting; N:~MANAGE~4, GOVERN~4.2: residual risk and incident protocols; ISO Cl.~9.1, 10: monitoring and corrective actions; A.8: information for interested parties \\
\hline
0.3 Risk Mgmt
& Cross
& N:~GOVERN (all 6): culture, policy, accountability; EU AI Act: AI Office and MSA oversight; ISO: PDCA cycle; Cl.~5: leadership; Cl.~9: audits \\
\hline
0.8 Ethics
& Cross
& N:~GOVERN~4, MAP~5: ethics and societal impact; EU AI Act Art.~10(2)(g), 10(3): bias and representativeness; ISO A.6: AI system lifecycle controls \\
\hline
0.9 Audit \& CI
& Cross
& N:~GOVERN~1, MEASURE: accountability and re-assessment; EU AI Act Arts.~16--17, 72: QMS and post-market monitoring; ISO Cl.~5, 9, 10: leadership, audits, improvement \\
\hline
\end{tabular}
\vspace{1mm}
\parbox{\linewidth}{\scriptsize N\,=\,NIST AI RMF; ISO\,=\,ISO/IEC 42001;
GPAI\,=\,General-Purpose AI obligations under the EU AI Act (Arts.~53--55);
L1--L4\,=\,Data, Model, Distribution, Application; Cross\,=\,cross-cutting.
Category order matches Figure~\ref{fig:llm_lifecycle_governance_stack_v2}.}
\end{table}
\FloatBarrier

Three features of this distribution are worth noting. The first is that the categories most directly shaping the system, namely Data Governance and Model Safety and Alignment, fall at the upstream end of the lifecycle, where decisions about training corpora, alignment strategy, and capability boundaries are made and where
the consequences of those decisions become difficult to reverse. The second is that the categories carrying the heaviest external accountability, namely Transparency and Documentation and Regulatory Compliance, cluster at the boundary between model development and distribution, where evidence produced inside the organisation is handed over to external regulators. The third is that Risk Management, Ethics and Responsible AI, and Audit and Continuous Improvement apply throughout the lifecycle rather than
attaching to any one layer; they operate as cross-cutting concerns that interact with every other category.

\subsection{Governance Handoffs}
\label{sec:gov-handoffs}

Three formal governance handoffs were identified at layer
boundaries, with the identification criteria stated in
Section~\ref{sec:methodology}. The \emph{first handoff}
(Layer~1$\to$2) transfers data governance responsibility from data engineering to model development. The data team must produce provenance records, bias assessments, consent documentation, and training data summaries (EU AI Act Art.~10, Art.~53(1)(d)); the model development team consumes these as the evidentiary foundation for model governance. No external enforcement mechanism verifies
completeness before consumption, with the consequence that if these artifacts are incomplete, model governance begins on uncertain foundations that downstream testing cannot fully remediate.

The \emph{second handoff} (Layer~2$\to$3) is the most consequential of the three, since at this boundary the provider formally releases the model and accountability transitions to the external regulatory chain. The model development team must transmit conformity declarations (EU AI Act Art.~47), technical documentation (Annex~IV), model cards, and, where applicable, GPAI transparency reports (Art.~53) to the distribution function. The EU AI Act's
role distinctions become legally operative at this boundary, and releasing without complete documentation propagates irrecoverably into the regulatory chain.

The \emph{third handoff} (Layer~3$\to$4) transfers primary
accountability from provider to deployer. The provider retains responsibility for post-market monitoring and incident cooperation (EU AI Act Arts.~72--73), while the deployer assumes accountability for human oversight (Art.~14), logging (Art.~12), and user disclosure (Art.~50). The deployer typically lacks visibility into
upstream development decisions, creating an information asymmetry between the entity bearing deployment-time accountability and the entity that made the foundational choices on which that accountability rests.

\subsection{Structural Asymmetry}
\label{sec:gov-asymmetry}

The governance mapping surfaces a notable structural property of the current regulatory landscape: governance obligations and evidence-production requirements concentrate at the deployment-facing end of the lifecycle, while the decisions that most fundamentally determine system security are made at the development-facing end, where regulatory visibility is lowest.

The asymmetry operates along two dimensions. At Layers~1--2, provisions are predominantly decision-shaping, governing what data to collect, what alignment strategy to pursue, and what capability boundaries to impose. These choices are difficult or impossible to reverse once training completes. At Layers~3--4, provisions shift
to evidence-producing, requiring documentation, conformity
processes, logging, incident reporting, and oversight
implementation. Quantitatively, the asymmetry is near-monotonic: of the thirty-three governance provisions mapped to lifecycle-bound categories in Table~\ref{tab:gov-mapping}, ten of the eleven provisions assigned to Layers~1--2 are decision-shaping, whereas all twenty-two provisions assigned to Layers~2/3 through~4 are
evidence-producing.

At the same time, the EU~AI Act places its heaviest enforcement at Layers~3--4, with fines up to~\EUR15~million or 3\% of global
turnover, while the most consequential decisions are made at Layers~1--2, where enforcement is absent. The practical consequence is that an organisation can satisfy every downstream requirement while having made unverified foundational decisions upstream. This misalignment has direct security implications, since data poisoning, alignment subversion, and capability-boundary violations
originate at Layers~1--2, and by the time systems reach
Layers~3--4 these compromises are already embedded in model
weights. The GPAI provisions (Arts.~53--55) partially bridge this gap by imposing transparency and adversarial-testing requirements on foundation-model providers, but they do not resolve it: the provisions remain evidence-producing rather than decision-shaping, and they do not subject upstream design choices to external
verification.

\section{Discussion}
\label{sec:comparison}

The model proposed in this paper defines 32 core lifecycle stages, 12 LLMOps sub-stages, and 9 governance categories. Together, these create 53 points where security and governance questions can be asked. The aim is not to claim that these are the only possible boundaries, but to make important differences visible, for example between packaging and signing, input validation and inference, or alignment and supervised fine-tuning. On \emph{architectural completeness}, the model combines end-to-end coverage from data sourcing through decommissioning with both an operational and a governance pillar as cross-cutting structures, a combination we did not identify among the frameworks surveyed in Table~\ref{tab:comparison}. On \emph{security-specific decomposition}, the 13~stages introduced here as distinct lifecycle stages fall into three categories: security-critical activities previously subsumed under broader categories (1.4, 1.5, 2.4, 2.5), lifecycle activities omitted entirely (1.6, 2.8, 2.9, 3.4, 3.5, 4.12), and emerging architectural paradigms (4.9, 4.10, 4.11). The \emph{governance mapping} also points to an important imbalance. Many of the most important security decisions are made early, during data and model development, but much of the formal evidence, documentation, and accountability appears later, around distribution and deployment. 

Three incident mappings illustrate this finer decomposition's analytical value. These examples should be read as illustrative mappings, not as empirical validation of the model. First, the 2024 Hugging Face serialization attacks~\cite{Casey_LargeScale_2024} map to Stages~2.7--3.2, a distinction between packaging, signing, and registry that two-phase models collapse into a single category, obscuring where the trust boundary was breached. Second, reward model poisoning attacks~\cite{Wang_RLHFPoison_2024} target Stage~2.4 (Alignment) specifically; frameworks that merge alignment with SFT cannot distinguish whether the poisoning vector is instruction data or preference labels, yielding different mitigations. Third, agentic exploitation attacks target Stage~4.11 (Agency and Permission Control) through tool-use protocols such as MCP, a stage no framework reviewed in Section~\ref{sec:background} treats separately.

The model also has several limitations. The most fundamental is complexity: 53~addressable units provide resolution for per-stage analysis but impose cognitive cost. For this reason the model can be adopted gradually using a three-tier adoption model: Tier~1 adopts the four-layer abstraction and three governance handoffs as an organizational orientation tool; Tier~2 adds the 13~additional stages as priority extensions to existing frameworks; Tier~3 adopts the full 32-stage decomposition with LLMOps pillar as organizational security maturity increases. The framework is intentionally detailed:  organisations that skip stages inherit the residual risk that the skipped stage was designed to address, and the framework makes this risk explicit at the point of omission.

The continuous feedback loop from Layer~4 to Layer~1 is not assigned its own stage but is governed by three distributed components: Stage~4.8 (Feedback Collection), Stage~1.1 (Data Sourcing, which consumes feedback as retraining data), and LLMOps sub-stage~L.7 (Human Feedback Integration, which orchestrates the loop); dedicated security analysis of the loop's properties is distributed across these units. Certain stages are LLM-specific (alignment, prompt engineering, agentic orchestration); extending to other foundation model types would require modifications. 

The governance mapping is interpretive. Other analysts may assign some provisions to neighbouring stages, especially broad provisions related to risk management or ethics. However, the main pattern is likely to remain: early lifecycle stages shape the system, while later stages produce much of the evidence used for oversight and accountability. The stage separation criterion is conservative by design and may cause under-decomposition where security-relevant distinctions exist but lack sufficient published research. Internal validity is bounded by the interpretive nature of stage derivation; external validity by focus on the current regulatory landscape; construct validity by the analytical rather than empirical stage derivation, which a companion study is designed to validate through systematic mapping of security concerns to the lifecycle structure.


\section{Conclusion}
\label{sec:conclusion}

This paper has presented a lifecycle model for LLM systems, designed to support security analysis, comprising 32~stages across four core pipeline layers organised by attack surface exposure, introducing 13~stages first identified here as distinct lifecycle stages, absent from all surveyed frameworks; a 12-stage LLMOps pillar elevated to a cross-cutting layer whose bidirectional relationship with the pipeline ensures dedicated security analysis of operational infrastructure; and a 9-category governance pillar synthesising the NIST AI~RMF, the EU~AI Act, and ISO/IEC~42001 into a lifecycle-mapped governance structure that identifies three governance handoffs and reveals a structural property of the current regulatory landscape, a misalignment between where governance evidence concentrates (Layers~3--4) and where the most consequential governance decisions are made (Layers~1--2). Future work will focus on a literature review to map known threats, attacks, and security controls to each stage of the proposed lifecycle. This would help validate the usefulness of the stage boundaries and support more practical per-stage security analysis. Additional priorities include practitioner validation, multi-modal extension, and upstream governance attestation mechanisms extending conformity-assessment logic to the development-facing stages where the most foundational security compromises originate.


\begin{credits}
\subsubsection{\discintname}
The authors have no competing interests to declare that are relevant to the content of this article.

\subsubsection{Acknowledgments.}
This work received partial funding from the European High-Performance Computing Joint Undertaking (JU) under Grant Agreement No. 101234269 for the \href{https://www.pharos-aifactory.eu}{Pharos AI Factory} project, as well as from the Greek Ministry of Digital Governance and Artificial Intelligence.
\end{credits}

\bibliographystyle{splncs04}
\bibliography{Lifecycle_of_LLMs3}

\end{document}